\documentclass[a4paper,aps,prd,twocolumn,nofootinbib]{revtex4}
\usepackage[mathscr]{eucal}
\usepackage[centertags]{amsmath}
\usepackage{dsfont}
\usepackage{yfonts} 
\usepackage{enumerate}
\usepackage[latin1]{inputenc}
\usepackage{amsfonts}
\usepackage{axodraw}
\usepackage{subfigure}
\usepackage{epsfig}
\usepackage{geometry}
\usepackage[dvips]{color}
\definecolor{Red}{cmyk}{0,1,1,0}
\usepackage{amssymb}
\usepackage{amscd}
\usepackage{mathrsfs}
\usepackage{latexsym}
\usepackage{amsmath}
\definecolor{White}{cmyk}{0,0,0,0}

\newcommand{\newc}{\newcommand}
\newc{\beq}{\begin{equation}}
\newc{\eeq}{\end{equation}}
\newc{\beqn}{\begin{eqnarray}}
\newc{\eeqn}{\end{eqnarray}}
\newc{\bsym}{\boldsymbol}
\newc{\lam}{\lambda}
\newc{\Lam}{\Lambda}
\newc{\kap}{\kappa}
\newc{\lra}{\leftrightarrow}
\newc{\nonum}{\nonumber}
\newc{\phm}{\phantom{-}}
\newc{\ra}{\rightarrow}
\DeclareMathAlphabet{\scr}{U}{rsfs}{m}{n}

\begin{document}
\pagestyle{plain}
\date{\today}
\title{{\bf{\Large A Simple Baryon Triality Model for 
Neutrino Masses\\~}}}

\author{Herbi K. Dreiner
}\!\!\email[E-mail:\ ]
{dreiner@th.physik.uni-bonn.de}
\author{Jong Soo Kim\,
}\email[E-mail:\ ]
{jsk@th.physik.uni-bonn.de}
\author{Marc Thormeier\,
}\email[E-mail:\ ]
{thor@th.physik.uni-bonn.de} 
\affiliation{~\\Physikalisches Institut der Universit\"at Bonn\\ 
        Nu\ss{}allee 12, 53115 Bonn, Germany\\~ }
\begin{abstract} \noindent
  We make a simple ansatz for the supersymmetric lepton-number
  violating Yukawa couplings, by relating them to the corresponding
  Higgs Yukawa couplings. This reduces the free B$_3$ parameters from
  36 to 6. We fit these parameters to solve the solar and atmospheric
  neutrino anomalies in terms of neutrino oscillations. The resulting
  couplings are consistent with the stringent low-energy bounds. We
  investigate the resulting LHC collider signals for a stau LSP
  scenario.
\end{abstract}
\maketitle
%
%
\section{Introduction\label{intro}}\noindent
The first experimental evidence for physics beyond the Standard Model
(SM) has been found in the neutrino sector \cite{neutrino-exp}. The
solar and atmospheric neutrino anomalies are best explained in terms
of oscillating massive neutrinos \cite{Wolfenstein:1977ue}, as opposed
to for example lepton flavour-violating interactions \cite{lfv}.
Assuming massive neutrinos, based on a three neutrino fit including
the recent MINOS \cite{Michael:2006rx} and the SK-II atmospheric data
\cite{sk-ii}, the corresponding neutrino mass and mixing parameters at
$1\sigma\,(3\sigma) $ C.L. are \cite{Gonzalez-Garcia:2004jd,GonzalezGarcia:2007ib}
\beqn 
\Delta m^2_{21} &=& 7.9{\small \begin{array}{l}{+0.27}\\[-1mm]{-0.28}\end{array}}
 {\small \left(\!\begin{array}{l}{+1.1}\\[-1mm]{-0.89}\end{array}\!\right)}
\times 10^{-5} {\rm eV}^2\,,\label{exp-neut1}\\ [2mm]
\left|\Delta m^2_{31}\right| &=& 2.6\pm 0.2\, (0.6)
\times 10^{-3}\,{\rm eV}^2\,, \label{exp-neut2}\\ [1mm]
 \theta_{12} &=& 33.7\pm 1.3   {\small \left(\!\begin{array}{l}{+4.3}\\
[-1mm]{-3.5}\end{array}\!\right)} ,\label{exp-neut3}\\ 
\theta_{23} &=&43.3 {\small \begin{array}{l}{+4.3}\\[-1mm]{-3.8}\end{array}}
 {\small \left(\!\begin{array}{l}{+9.8}\\[-1mm]{-8.8}\end{array}\!\right)}
\,, \label{exp-neut4}\\
\theta_{13}&=&0 {\small
  \begin{array}{l}{+5.2}\\[-1mm]{-0.0}\end{array}}
{\small\left(\!\begin{array}{l}{+11.5}\\[-1mm]{-0.0}\end{array}\!\right)}
\,.\label{exp-neut5} 
\eeqn 
The angles are given in degrees.  The most widely discussed extensions
of the SM to include massive neutrinos involve the see-saw mechanism
\cite{see-saw}. These require right-handed neutrinos, as well as a
new, typically very large Majorana mass-scale. The see-saw mechanism
can also be incorporated into the minimal supersymmetric SM (MSSM)
\cite{susy-review}, now requiring right-handed neutrino superfields,
as well as the additional high mass scale.

\medskip

However, within supersymmetry there is another, in our opinion
simpler, possibility to include massive neutrinos, namely via \textit{
renormalizable} lepton-number violating terms, $W_{\not L_i}$, in the
superpotential \cite{Hall:1983id},
\begin{eqnarray}
W_{\not L_i}&=& \lam_{ijk} L_i  L_j {\bar E}_k+ 
\lam^\prime_{ijk} L_i  Q_j {\bar D}_k 
+ \kap_i L_i H_u\,,~~~~
\label{lep-rpv-superpot} 
\end{eqnarray}
\begin{eqnarray}
W_{H_d} &=& \hspace{-0.18cm}h^E_{ij} L_i H_d 
{\bar E}_j + h^D_{ij}~  Q_i  H_d {\bar D}_j 
+ \mu~  H_d H_u\,,\label{Hd-superpot}
\end{eqnarray}
where we have employed the conventional notation for the superfields
\cite{Allanach:2003eb}. For later use, we have also included the
superpotential terms involving the down-like Higgs superfield. The
terms in $W_{\not L_i}$ violate $R$-parity (a $\mathds{Z}_2$-symmetry)
as well as proton hexality \cite{Dreiner:2005rd,Dreiner:2007vp} (a
$\mathds{Z}_6$-symmetry), but conserve baryon triality ($B_3$, a
$\mathds{Z}_3$-symmetry, sometimes also misleadingly called baryon
parity) \cite{Ibanez:1991pr,Dreiner:1997uz,Barbier:2004ez}. The
Majorana neutrino masses are generated via tree-level mixing with the
neutralinos, as well as via radiative corrections
\cite{Hall:1983id,Hirsch:2000ef,Diaz:2003as,Grossman,davidson,hempfling,borzumati}.
There is an implicit see-saw mechanism in the neutralino-neutrino
sector: $\kap_i^2/M_{1/2} $, but with a much smaller hierarchy of mass
scales. Furthermore, no new fields or mass scales are required.

\medskip

Within a baryon triality supergravity model the largest neutrino mass is
naturally small \cite{Allanach:2003eb}. For universal soft breaking
terms, the mixing, $\kap_i$, with the neutralinos is zero at the
unification scale. It is subsequently generated at the order of a few
MeV via renormalization group equations. It is thus proportional to
the product of a (small) down-like Higgs Yukawa coupling (for example
of the bottom quark or the tau lepton), a (small) baryon triality
coupling and the Higgs mixing parameter $\mu$
\cite{Nardi,deCarlos:1996du,Allanach:1999mh,Allanach:2003eb}. The 
lighter neutrino masses are generated via radiative corrections, and
are naturally further suppressed.

\medskip

There are $9+27+3=39$ lepton-number violating (complex) parameters in
the superpotential $W_{\not L}$.  There are also 39 corresponding
soft-supersymmetry breaking parameters, which in principle are
independent, but are usually related to those of $W_{\not L}$ via
universal soft-supersymmetry breaking \cite{Soni:1983rm}. In a
top-down approach, \textit{e.g.} based on the Froggatt-Nielsen
mechanism \cite{FN-1979}, one can attempt to predict the order of
magnitude of all superpotential parameters, \textit{i.e.} $W_{\not
L_p}$ together with the Higgs Yukawa couplings, based on a
spontaneously broken gauge symmetry,
Ref.~\cite{Dreiner:2003hw,frogg-niels} and references therein. See
also Refs.~\cite{hempfling,mnrv,borzumati,models}. 

\medskip

In this letter, we instead propose a baryon triality model of neutrino
masses, based on a simple phenomenological ansatz, which relates the
Higgs superpotential parameters to those that violate lepton-number.
The justification for this is that the down-like Higgs doublet
superfield and the lepton-doublet superfields have identical Standard
Model gauge quantum numbers. We make no assumption about the possible
underlying theory at the unification scale. This ansatz dramatically
reduces the number of free parameters. If experimentally confirmed it
would give a clear indication on how to construct the more fundamental
unified theory.

\medskip

In the literature there are other simple ans\"atze
\cite{Drees:1997id,Chun:1998gp,Choi:1999tq}, the most common and also
the most similar to ours is pure bi-linear lepton-number violation,
\textit{i.e.} $\lam_{ijk}=\lam'_{ijk}=0$, and $\kap_i\not=0$. For this
there is an extensive literature, see for example
\cite{hempfling,Hirsch:2000ef,Diaz:2003as,mnrv,Kaplan:1999ds,Abada:2001zh}
and references therein. We discuss how our ansatz differs from the
bi-linear case in Sect.~\ref{sub-sec}.

\medskip

A special feature of baryon triality models for the neutrino masses, is
that they lead to other observable effects at colliders and can thus
be tested \cite{Mukhopadhyaya:1998xj,Datta:1999xq,Choi:1999tq,Porod:2000hv,Datta:2000ci,Chun:2002rh,Barger:2001xe,Hirsch:2002ys,Hirsch:2002tq,Magro:2003zb,Hirsch:2003fe,deCampos:2005ri,Das:2003pe,Jung:2004rd,Datta:2005yt,Das:2005mr,Datta:2006ak}.
In the case of pure tri-linear couplings ($\kap_i=0$), a fit to the
neutrino data, Eqs.~(\ref{exp-neut1})-(\ref{exp-neut5}), leads to values
in the range $\lam_{ijk},\,\lam'_{ijk}\sim 10^{-5} - 10^{-4}$
\cite{Rakshit:1998kd,Abada:1999ai,Abada:2000xr,Abada:2002ju}. These 
couplings are very small, in particular, too small for the resonant
production of supersymmetric particles \cite{resonant}. However, they
do lead to the decay of the lightest supersymmetric particle in the
detector, possibly with a detached vertex. This model can be confirmed
by measuring the branching ratios of the various lightest
supersymmetric particle (LSP) decays and thereby measuring the
couplings. However, several points have in my opinion been missed in
the literature. In the case of a pure fit,
\textit{i.e.} not a model, it is possible to have larger couplings, 
which do not contribute to the neutrino masses, or which are not
required for the fit. In this case, the LSP decay which dominates the
collider signals will be completely independent of the neutrino
sector. Thus pure fit models can only be tested if the neutrino mass
parameters dominate the $B_3$ sector. We consider here a complete
model, where the fit to the neutrino data fixes \textit{all} the $B_3$
parameters. Second, it has hitherto been assumed, that the LSP is the
lightest neutralino. We go beyond this and also consider a scalar tau
LSP \cite{Allanach:2003eb,Allanach:2006st}.

\medskip

Our analysis is structured as follows: In Sect.~\ref{simplerpv}, we
present our model in detail. We then briefly review the neutrino
masses in baryon triality models, Sect.~\ref{numass}. In
Sect.~\ref{model}, we estimate the values of the free parameters which
result in acceptable neutrino masses. In Sect.~\ref{numer-eval}, we
numerically evaluate the new parameters in our model, such that the
neutrino masses and mixing angles fall in the required experimental
ranges, \textit{cf} Eqs.~(\ref{exp-neut1})-(\ref{exp-neut5}). In order
to obtain at least two non-vanishing neutrino masses, we must violate
at least two lepton numbers. This typically leads to significantly
stricter bounds on the products of couplings
\cite{Davidson:1993qk,Smirnov:1996bg,Allanach:1999ic,Dreiner:2006gu}. 
In Sect.~\ref{bounds}, we investigate, whether our model is
consistent with these bounds. In Sect.~\ref{collider}, we discuss
possible future tests of the ansatz at colliders, in particular the
LHC.  In Sect.~\ref{concl} we conclude.

\section{Simple B$_3$-Model\label{simplerpv}}\noindent 
%
\noindent
In the MSSM, the lepton doublet superfields $L_i$ and the down-type
Higgs superfield $H_d$ have the same gauge quantum numbers. They are
distinguished through a discrete symmetry: lepton number. However, in
the case of baryon triality, lepton number is violated and not well
defined. In the most general baryon triality superpotential with the
MSSM superfields, $H_d$ and $L_i$ have exactly corresponding terms in
the superpotential, as can be seen in Eqs.~(\ref{lep-rpv-superpot})
and (\ref{Hd-superpot}).  We take this correspondence to motivate the
following simple ansatz for the Yukawa coupling constants
\begin{eqnarray}
\lam_{ijk}&\equiv&\ell_{i}\cdot h^E_{jk}~-~\ell_{j}\cdot h^E_{ik}
\label{ansatz1}\,,
\\
\lam^\prime_{ijk}&\equiv&\ell^\prime_{i}\cdot h^D_{jk}
\label{ansatz2}\,, \\
\kappa_i&\equiv&c_i\cdot\mu \,. \label{ansatz3}
\end{eqnarray}
Here, $\ell_i,\,\ell^\prime_i$ are $c$-numbers. Eq.~(\ref{ansatz1})
has the required form to maintain the anti-symmetry of the
$\lam_{ijk}$ in the first two indices. The ansatz for the dimensionful
mixing terms, Eq.~(\ref{ansatz3}), is no simplification and we retain
the $\kap_i$ as free parameters.  Given the ansatz of
Eqs.~(\ref{ansatz1}),\,(\ref{ansatz2}), and assuming we know the
Higgs-Yukawa coupling constants (leading to the SM fermion mass
matrices), then the 36 couplings $\lam_{ijk}$, $\lam^\prime_{ijk} $,
are parameterized in terms of the six numbers $\ell_i,\,\ell^\prime
_j$.

\medskip

Since $L_i$ and $H_d$ have the same {\it gauge} quantum numbers, our
ansatz in Eqs.~(\ref{ansatz1}),\,(\ref{ansatz2}) is given in the $SU
(2)\times U(1)$ {\it current}-eigenstate basis. Thus when computing
neutrino masses and comparing the required Yukawa coupling constants
to low-energy bounds, we must rotate to the mass-eigenstate basis
\cite{Dreiner:1991pe,Agashe}. This requires a bi-unitary transformation
in generation space. We shall denote the transformation of the
left-handed and right-handed fermions (not superfields), respectively
by
\begin{eqnarray}
\boldsymbol{e_L}\!\!&=&\!\!\boldsymbol{V_{e}} 
\,\boldsymbol{e_L^{\,\prime}}
\,,\quad\boldsymbol{d_L}=\boldsymbol{V_d}
\, \boldsymbol{d_L^{\,\prime}}\,, \quad \boldsymbol{u_L}=
\boldsymbol{V_u} \,
\boldsymbol {u_L^{\,\prime}}\,, \\[1.5mm]
\boldsymbol{e_R}\!\!&=&\!\!\boldsymbol{U_e} \,
\boldsymbol{e_R^{\,\prime}}
\,,\;\;
\boldsymbol{d_R}=\boldsymbol{U_d} \,\boldsymbol{d_R^{\,\prime}}\,,\;\;
\boldsymbol{u_R}=\boldsymbol{U_u} \,\boldsymbol{u_R^{\,\prime}}\,,
\end{eqnarray}
where ${\boldsymbol{V}}_{\boldsymbol{e,d,u}},\,{\boldsymbol{U_{e,d,u}}}$ are $3\times3$ matrices
in generation space and the prime denotes the mass-eigenstates. We
have combined the charged lepton and quark states into three-component
vectors in generation space, {\it e.g.}  ${\boldsymbol{e_L}}\equiv(e_L,\mu_L,
\tau_L)$. We rotate the sfermion partners by the same matrices in
flavour space. By construction, these transformations diagonalize the
SM Yukawa coupling matrices
\begin{eqnarray}
\boldsymbol{U_{e}}^{\dagger}\cdot(\boldsymbol{h^{{E}}})^T\cdot 
\boldsymbol{V_{e}}&=&\frac{\sqrt 2}{\upsilon_{d}}~\text{{\bf{diag}}}
(m_{e},m_{\mu}, m_{\tau})\,,\; \label{diag1}\\
\boldsymbol{U_{d}}^{\dagger}\cdot(\boldsymbol{h^{{D}}})^{T} 
\cdot\boldsymbol{V_{d}}&=&\frac{\sqrt 2}{\upsilon_{d}}~\text{{\bf{diag}}}
(m_{d},m_{s},m_{b})\,,\label{diag2}\\
\boldsymbol{U_{u}}^{\dagger}\cdot(\boldsymbol{h^{{U}}})^{T}\cdot 
\boldsymbol{V_{u}}&=&\frac{\sqrt 2}{\upsilon_{u}}~\text{{\bf{diag}}}
(m_{u},m_{c}, m_{t})\,,
\end{eqnarray}
where the normalization of the Higgs vacuum expectation value is $v=
\sqrt{|\upsilon_u|^2+ |\upsilon_d|^2}=246\,\text{GeV}$ and 
$\tan\beta\equiv \upsilon_u/\upsilon_d$.

\medskip

In the following, we assume that the charged lepton mass- and
weak-eigenstates are the same. The corresponding charged lepton
rotation matrices are then given by $\boldsymbol{V_e}=\boldsymbol{U_e}
=\boldsymbol{1}\!\!\boldsymbol{1}$. Thus in our ansatz, in the
leptonic sector the mixing takes place entirely in the neutrino
sector, \textit{cf.} the discussion in
Refs.~\cite{Allanach:1999ic,Agashe}. Using Eqs.~(\ref{ansatz1}), and
(\ref{diag1}), the $LL{\bar E}$ couplings can then be expressed in
terms of the lepton masses and the three parameters $\ell_i$
\begin{equation}
\lam_{ijk}=\ell_i \frac{\displaystyle \sqrt{2}m_{e^j}}
{\displaystyle \upsilon_d}\delta_{jk}-\ell_j\frac{\displaystyle \sqrt{2}m_{e^i}}
{\displaystyle \upsilon_d}\delta_{ik}\,,
\label{mass-ansatz1}
\end{equation}
where $m_{e^j}\equiv(m_e,m_\mu,m_\tau)^i$. Explicitly the couplings are
given in Table~\ref{table-LLE}, as a function of the free parameters
$\ell_i$. As an example, we have also given numerical coefficients in
the case where $\tan\beta=10$, which fixes $\upsilon_d=24.5\,\mathrm{GeV}$.
Overall, of course, all couplings are proportional to the free
parameters $\ell_i$.

\begin{table}{\begin{tabular}{|c|c|c|}\hline
\;Coupling\; & Model Value & Numerical Value \\
&& ($\tan\beta=10$) \\ \hline
$\lam_{121}$ & $\;-\ell_2 {\sqrt{2}m_e}/{\upsilon_d}\;$ & $\;-2.9\cdot10^{-5}
\cdot\ell_2\;$\\ \hline
$\lam_{122}$ & $\phm\ell_1 {\sqrt{2}m_\mu}/{\upsilon_d}$ & $\phm6.1\cdot 10^{-3}\cdot\ell_1$\\ \hline
$\lam_{123}$ & 0 & 0 \\ \hline
$\lam_{131}$ & $-\ell_3 {\sqrt{2}m_e}/{\upsilon_d}$ & $-2.9\cdot10^{-5}\cdot\ell_3$\\ \hline
$\lam_{132}$ & 0 & 0 \\ \hline
$\lam_{133}$ & $\phm\ell_1 {\sqrt{2}m_\tau}/{\upsilon_d}$ & $\phm1.0\cdot10^{-1}\cdot\ell_1$\\ \hline
$\lam_{231}$ & 0 & 0 \\ \hline
$\lam_{232}$ & $-\ell_3 {\sqrt{2}m_\mu}/{\upsilon_d}$ & $-6.1\cdot10^{-3}\cdot\ell_3$\\ \hline
$\lam_{233}$ & $\phm\ell_2 {\sqrt{2}m_\tau}/{\upsilon_d}$ & $\phm1.0\cdot10^{-1}\cdot\ell_2$\\ \hline
\end{tabular}}
\caption{Predictions for the $LL\bar E$ in our ansatz as a function 
of the free parameters $\ell_i$. \label{table-LLE}}
\end{table}

\medskip

We thus have some very specific predictions for the $LL\bar E$
couplings in our model.
\begin{eqnarray}
\lam_{123}=\lam_{132}=\lam_{231}&=&0 \label{zero}\\[2mm]
\frac{\lam_{121}}{\lam_{233}} = -\frac{m_e}{m_\tau},\quad\;
\frac{\lam_{122}}{\lam_{133}} = \frac{m_\mu}{m_\tau},\quad\;
\frac{\lam_{131}}{\lam_{232}} &=& \frac{m_e}{m_\mu}\,.\; \;\;\;
\label{zero+1}
\end{eqnarray}
Besides the vanishing couplings, we would thus expect the couplings to
satisfy the strict constraints
\begin{eqnarray}
\lam_{121}&<&2.0\cdot10^{-5}\,\frac{m_{\tilde\tau_R}}{100\,\mathrm{GeV}}\,,\\[2mm]
\lam_{122}&<&3.6\cdot10^{-4}\,\sqrt{\frac{m_{\tilde\tau}}{100\,\mathrm{GeV}}}\,,\\[2mm]
\lam_{131}&<&3.4\cdot10^{-4}\,\frac{m_{\tilde\mu_R}}{100\,\mathrm{GeV}}\,,
\end{eqnarray}
where we have implemented the low-energy bounds in
\cite{Allanach:1999ic} for $\lam_{133},\,\lam_{233},\,\lam_{232}$ and
inserted the PDG lepton masses \cite{Yao:2006px}.

\medskip

Throughout we consider our model only at the weak scale. In principle
it should be embedded in a unified model at the grand unified scale or
above \cite{frogg-niels,Dreiner:2003hw}. In that case, the predictions
in Eqs.~(\ref{zero}),\,(\ref{zero+1}) would be modified by
renormalization group effects. In particular the couplings in
Eq.~(\ref{zero}) would get non-zero contributions
\cite{Allanach:1999mh}, which however are extremely small, as they are
proportional to the product of three non-zero $LL\bar E$ couplings.

\medskip

When expanding the $LL{\bar E}$ term in the Lagrangian into its
mass-eigenstate components, we obtain (summation over generation
indices implied)
\begin{eqnarray}
\mathscr{L}_{LL\bar{E}}&=&\left[-\lam_{ijk}\tilde e_{R}^{k*}\bar \nu^{ic} P_L e^j-
\lam_{ijk}\tilde e_{L}^j\bar e^k P_L \nu^i\right.\nonum\\
&&-\lam_{ijk}\tilde \nu_{L}^i\bar e^k P_L e^j\Bigr] +h.c.\,.
\end{eqnarray}
\medskip
Next we consider the $LQ{\bar D}$ term in the Lagrangian. Expanding
out the $SU(2)_L$ doublet superfields, we obtain
\begin{equation}
\mathscr{L}_{LQ\bar{D}}=\lam_{ijk}^\prime{{N}_{Li}} {{D}_{Lj}} 
\bar{D}_{Rk}-\lam_{ijk}^\prime {{E}_{Li}} {{U}_{Lj}}
\bar{D}_{Rk}\,.
\label{lqd}
\end{equation}
Rotating the quark superfields in the first term into the superfield
basis where the quarks are in the mass-eigenstate and using
Eqs.~(\ref{ansatz2}) and (\ref{diag1}) we obtain
\begin{eqnarray}
\lam_{ijk}^\prime {{N}_{Li}} {{D}_{Lj}} \bar{D}_{Rk}
&=&\lam_{ijk}^\prime\,[{\boldsymbol{U}}_{\boldsymbol{d}}^{\dagger}]_{rk}
{[\boldsymbol{V}_{\boldsymbol{d}}]}_{js} 
{{{N}_{Li}}} {{D}_{Ls}}^\prime {{\bar D}}_
{Rr}^\prime \nonum\\
&=&\ell_{i}^\prime\,[\boldsymbol{U}_{\boldsymbol{d}}^{\dagger}
 ({\boldsymbol{h}^{\boldsymbol{D}}})^{T}
{\boldsymbol{V}_{\boldsymbol{d}}}]_{rs} N_{iL} D_{Ls}^\prime
{{\bar D}}_{Rr}^\prime \nonum \\
&=& \frac{\sqrt{2}\,\ell_{i}^\prime}{\upsilon_d}\, m_{d^r}\,\delta_{rs}\,
 N_{iL} D_{Ls}^\prime
{{\bar D}}_{Rr}^\prime.\label{one} 
\end{eqnarray}
For the second term in Eq.(\ref{lqd}), we obtain analogously
\begin{eqnarray}
\lam_{ijk}^\prime{{E}_{Li}}{U}_{Lj}{\bar{D}}_{Rk} \nonum \\
=\ell_{i}^\prime ~[{\boldsymbol{U}_{\boldsymbol{d}}}^\dagger\hspace{-0.2cm}
&&\!\!\!\!
({\boldsymbol{h}^{\boldsymbol{D}}})^{T} \boldsymbol{V}_{\boldsymbol{d}}
]_{rt} ~ [{\boldsymbol{V}_{\boldsymbol{d}}}^\dagger\boldsymbol{V_u}]_{ts} 
\nonum  {E}_{Li}^\prime
{{U}_{Ls}}^\prime{{\bar{D}}}_{Rr}^\prime\nonum \\
=\frac{\sqrt{2}\,\ell'_i}{\upsilon_d}&&\hspace{-0.25cm}
m_{d^r}\, [\boldsymbol{V^\dagger}]_{sr}\; {{E}_{Li}}^\prime
{{U}_{Ls}}^\prime{{\bar{D}}}_{Rr}^\prime \,, \label{two}
\end{eqnarray}
where $\boldsymbol{V_{CKM}}=\boldsymbol{V_u}^{\dagger}\boldsymbol{V_d}$ is
the Cabibbo-Kobayashi-Maskawa matrix.  Combining Eqs.~(\ref{one}) and (\ref{two}), we can
then write the Lagrangian for the $LQ{\bar D}$ interactions in the
mass-eigenstate basis and expanded in superfield
components\footnote{The primes, denoting the mass-eigenstates, are
  omitted in the following.}
\begin{eqnarray}
\mathscr{L}_{LQ\bar{D}}\nonum \\ 
&&\hspace{-1.2cm}= \left[ -\tilde\lam_{ijk}^{\prime}\left(\tilde\nu^i_L\bar 
d^j P_L 
d^k+\tilde d_L^j \bar d^kP_L\nu^i+\tilde d_R^{j*}\bar\nu^{ic}P_Ld^k\right)
\right.
\nonum\\
&&\hspace{-1.2cm}+\;\left.\tilde\lam_{ijk}^{\prime}V_{rj}^*\left(\tilde e_L^i
\bar d^kP_Lu^r+ \tilde u_L^r \bar d^k P_Le^i+\tilde d_R^{k*}\bar 
e^{ic}P_Lu^r\right)\right]\nonum\\
&&\hspace{-1.2cm}+\;h.c.\,,
\end{eqnarray}
with the coupling defined by
\begin{equation}
\tilde\lam_{ijk}^{\prime}\equiv \ell_{i}^\prime\frac{\sqrt{2}\,m_{d_k}}
{\upsilon_d}\,\delta_{jk}\,.
\label{mass-ansatz2}
\end{equation}
We also introduce the notation
\begin{equation}
\tilde{\tilde\lam}_{ijk}^{\prime}\equiv \ell_{i}^\prime\frac{\sqrt{2}\,m_{d_k}}
{\upsilon_d}\,[\boldsymbol{V}]_{jk}\,.
\label{mass-ansatz2a}
\end{equation}
Note that the (s)neutrino interactions are flavour diagonal in the
down-(s)quarks, whereas the charged (s)lepton interactions involve
generation off-diagonal (s)quark interactions \cite{Agashe}. We have
given an estimate of the couplings $\tilde\lam^\prime$ and $\tilde{
\tilde\lam} ^\prime$, modulo the $\ell_i$ in Table~\ref{table-LQD}.
We have again assumed $\tan\beta=10$. Furthermore, we have taken the
central PDG values for the quark masses: $m_d=6\,\mathrm{MeV}$, $m_s=
103\,\mathrm {MeV}$, $m_b=4.2\,\mathrm{GeV}$, and the central values of
the global PDG fit for the CKM matrix entries (2 significant figures)
\cite{Yao:2006px}
\begin{equation}
\boldsymbol{V_{CKM}}=\left(
\begin{array}{ccc}
0.97 & \;0.23\; & 0.0040  \\
0.23 \;& 0.97&  0.042\\
0.0081& 0.042&1.0
\end{array} \right)\,.
\end{equation}
As can be seen from Table~\ref{table-LQD}, we thus have also simple
predictions for the $\tilde{\tilde\lam}^\prime$ couplings in terms of
quark masses and $\boldsymbol{V_{CKM}}$ entries and independent of
$\tan\beta$, 
\begin{equation}
\frac{\tilde{\tilde\lam}^\prime_{ijk}}{\tilde{\tilde\lam}^\prime_{ijl}}=
\frac{m_{d^k}V_{jk}}{m_{d^l}V_{jl}}
\,,\qquad
\frac{\tilde{\tilde\lam}^\prime_{ijk}}{\tilde{\tilde\lam}^\prime_{ilk}}=
\frac{V_{jk}}{V_{lk}}
\end{equation}

We can use Eqs.~(\ref{mass-ansatz1}), (\ref{mass-ansatz2})
(\ref{mass-ansatz2a}) to translate between our parameters $\,\ell_i,\,
\ell_i'$ and the $B_3$ couplings, where in the latter case, care must
be taken to include the CKM-mixing for the charged (s)lepton
interactions.

\begin{table}{\begin{tabular}{|c|c|c|}\hline
\;Index\; & $\tilde\lam^\prime/\ell_i^\prime$ &  $\tilde{\tilde\lam}^\prime/
\ell_i^\prime$ \\\hline
(i11) & $\;3.5\cdot 10^{-4}\;$&$\;3.4\cdot 10^{-4}\;$ \\ \hline
(i12) & 0 &$\;1.4\cdot 10^{-3}\;$ \\ \hline
(i13) & 0 &$\;9.7\cdot 10^{-4}\;$ \\ \hline
(i21) & 0 &$\;8.0\cdot 10^{-5}\;$ \\ \hline
(i22) & $6.0\cdot 10^{-3}$&$\;5.8\cdot 10^{-3}\;$ \\ \hline  
(i23) & 0 &$\;1.0\cdot 10^{-2}\;$\\ \hline
(i31) & 0 &$\;2.8\cdot 10^{-6}\;$\\ \hline
(i32) & 0 &$\;2.5\cdot 10^{-4}\;$\\ \hline
(i33) & $0.24$ & $0.24$ \\ \hline
\end{tabular}}
\caption{Predictions for the $LQ\bar D$ in our ansatz as a function 
of the free parameters $\ell_i^\prime$. In the right column we have assumed
$\tan\beta=10$.  \label{table-LQD}}
\end{table}

\medskip

It is the purpose of this letter to investigate whether with this
reduced freedom in the $B_3$ sector, we can still obtain neutrino
masses and mixings, which are, first of all, consistent with
Eqs.~(\ref{exp-neut1})-(\ref{exp-neut5}) and second, where the
resulting coupling constants are consistent with the existing
low-energy bounds. In Sect.~\ref{collider}, we then study possible
observable consequences of the absolute values of the couplings as
well as of the relative values.

\section{Other Ans\"atze \label{sub-sec}}

In Ref.~\cite{Chun:1998gp,Choi:1999tq,Jung:2004rd} the hierarchy in
the SM Higgs Yukawa couplings was taken to motivate a similar
hierarchy in the $LQ\bar D$ (and separately in the $LL\bar E$
couplings). The authors restrict themselves to the couplings
$\lam^\prime_ {i33}$ and $\lam_{i33}$. We extend this interesting work
in several respects. We include the most recent neutrino data in our
fit; Eqs.~(\ref{exp-neut1})-(\ref{exp-neut5}). Furthermore, we include
the CKM mixing in our ansatz, we thus have a prediction for the full
range of the couplings. This is particularly important for the
observable consequences of the model, \textit{i.e.} the LSP decays. We
also do combined fits including all the couplings, \textit{i.e.} the
$\lam$ and the $\lam^\prime$ couplings and also the $\kap_i$, our
Models \textbf{I} and \textbf{II} below in Sect \ref{numer-eval}.

The most widely considered simple ansatz are B$_3$ models, where
$\lam_{ijk}=\lam^\prime_{ijk}=0$ and $\kap_i\not=0$, often denoted
bi-linear R-parity violation. This clearly has only three free
parameters compared to the six or nine, in our models below. In order
to compare the two ans\"atze in more detail, we combine the fields
$\mathcal{L}_\alpha=(\mathcal{L}_0,\,\mathcal{L}_i) =(H_d ,\,L_i)$,
where $\alpha=0,\dots,3$ and $i=1,2,3$.  The bi-linear $R$-parity
violating superpotential is then
\begin{equation}
W= h^E_{ij}~ \mathcal{L}_i  \mathcal{L}_0
{\bar E}_j + h^D_{ij}~  Q_i  \mathcal{L}_0 {\bar D}_j 
+ \mu~  \mathcal{L}_0  H_u+ \kap_i L_i H_u\,,\label{2-superpot}
\end{equation}
We can now make a field redefinition 
\begin{equation}
\mathcal{L}\ra{\cal L}^\prime=\boldsymbol{R}\mathcal{L}, 
\end{equation}
such that the bi-linear lepton-number violating terms are eliminated
from the superpotential. The explicit form for $\boldsymbol{R}$ is
given in Ref.~\cite{Allanach:2003eb,Dreiner:2003hw,Diaz:2003as}. We
then obtain the superpotential
\begin{eqnarray}
\tilde W&=& h^E_{ij}~ [\boldsymbol{R}]_{i\alpha}[\boldsymbol{R}]_{0\beta}\,
\mathcal{L}_\alpha \mathcal{L}_\beta {\bar E}_j \nonumber \\
&+& h^D_{ij}~ [\boldsymbol{R}]_{0\alpha} Q_i \mathcal{L}_\alpha 
{\bar D}_j + \tilde\mu~ \mathcal{L}_0 H_u\,.\label{2a-superpot}
\end{eqnarray}
The transformed parameters (denoted by a tilde) are then given by
\begin{eqnarray}
\tilde\mu &=& \mu [\boldsymbol{R}]_{00} + \kap_i [\boldsymbol{R}]_{i0}\\[1mm]
{\tilde h}^D_{ij} &=& h^D_{ij} [\boldsymbol{R}]_{00} \\
{\tilde h}^E_{ij} &=& h^E_{lj} \biggl\{
[\boldsymbol{R}]_{li} [\boldsymbol{R}]_{00}-  
[\boldsymbol{R}]_{l0} [\boldsymbol{R}]_{0i}\biggr\}\label{tilde-hE}\\
{\tilde\lam}^\prime_{ijk} &=& h^D_{jk} [\boldsymbol{R}]_{0i} = 
\frac{ [\boldsymbol{R}]_{0i}}{[\boldsymbol{R}]_{00}} \tilde h^D_{jk}\label{tilde-hD} \\
{\tilde\lam}_{ijk} &=& 
h^E_{lk} \biggl\{[\boldsymbol{R}]_{lj} [\boldsymbol{R}]_{0i}
-[\boldsymbol{R}]_{li} [\boldsymbol{R}]_{0j}\biggr\}
\nonumber\\
&=& [\boldsymbol{R}]_{0i} \Bigl(h^E_{lk}
[\boldsymbol{R}]_{lj} \Bigr)-
[\boldsymbol{R}]_{0j} \Bigl([\boldsymbol{R}]_{li} h^E_{lk} \Bigr)
\,.\label{tilde-LLE}
\end{eqnarray}
In the last equation we have included the parentheses to emphasize the
sum over $l$. Note that not only are the tri-linear couplings $\lam_
{ijk},\,\lam^\prime_ {ijk}$ generated, but also the Higgs Yukawa
couplings are modified
\begin{equation}
h^{E,D}_{ij} \longrightarrow {\tilde h}^{E,D}_{ij}\,.
\end{equation}
{}From Eq.~(\ref{tilde-hD}) we see that 
\begin{eqnarray}
\lam^\prime_{ijk}&=&\ell^\prime_i \,
{\tilde h}^D_{jk}\,, \quad \mathrm{with}\quad \ell^\prime_i=\frac{ 
[\boldsymbol{R}]_{0i}}{[\boldsymbol{R}]_{00}}=\frac{\kap_i}{\mu}\,.
\end{eqnarray}
In the last equation, we have employed the explicit form of the matrix
$\mathbf{R}$. Next we would like to show that
\begin{eqnarray}
\tilde\lam_{ijk}&=&\ell_i {\tilde h}^E_{jk} - \ell_j {\tilde h}^E_{ik}\,.
\label{prop}
\end{eqnarray}
For this we insert ${\tilde h}^E$ from Eq.~(\ref{tilde-hE}) and factor $\ell_{i,j}$
\begin{eqnarray}
\tilde\lam_{ijk}&=& 
\ell_j [\boldsymbol{R}]_{00}\Bigl(h^E_{lk}[\boldsymbol{R}]_{li}\Bigr) 
- \ell_i [\boldsymbol{R}]_{00} \Bigl( h^E_{lk}[\boldsymbol{R}]_{lj} \Bigr)\nonum\\
&&\hspace{-0.5cm}
+\ell_j [\boldsymbol{R}]_{0i}\Bigl(h^E_{lk}[\boldsymbol{R}]_{l0}\Bigr)
-\ell_i [\boldsymbol{R}]_{0j}\Bigl(h^E_{lk}[\boldsymbol{R}]_{l0} \Bigr)
\label{deriv-2}
\end{eqnarray}
If we now set
\begin{eqnarray}
\ell_i=\frac{[\boldsymbol{R}]_{0i}}{[\boldsymbol{R}]_{00}}\,\qquad \mathrm{and}
\qquad \ell_j=\frac{[\boldsymbol{R}]_{0j}}{[\boldsymbol{R}]_{00}}\,.
\end{eqnarray}
the last two terms in Eq.~(\ref{deriv-2}) cancel and the first two
terms agree with Eq.~(\ref{tilde-LLE}). In particular, we see that B$_3$
models with only bi-linear terms are a special case of our ansatz with
$\ell_i=\ell^\prime_i$.

\begin{center}
\begin{figure}
\hspace*{-10mm}
\includegraphics[scale=0.8]{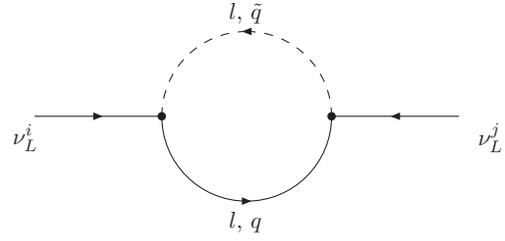}
\caption[Fig.]{\label{neutrinoloops}\footnotesize The radiative 
slepton-lepton and squark-quark contribution to the neutrino mass.}
\end{figure}
\end{center}

%
%
\section{B$_3$ Neutrino Masses \label{numass}}
\noindent
The general, lepton number violating superpotential in a $B_3$-model
is given in Eq.(\ref{lep-rpv-superpot}). As stated in the
introduction, due to the $\kap_i$, the neutrinos mix with the Higgsino
components of the neutralinos, resulting in one massive neutrino at
tree-level. The resulting mass matrix is perturbatively given
by\footnote{As in \cite{Grossman}, we assume here a rotation into the
basis where the sneutrino vacuum expectation values vanish.  In
principle this requires a detailed minimization of the scalar
potential as discussed for example in
Ref.~\cite{hempfling,Hirsch:2000ef,Allanach:2003eb}. This is well
beyond the scope of this paper.} \cite{Grossman,davidson,rakshit}
\begin{eqnarray}
({m^{\text{tree}}_{\nu}})_{ij}&=&\frac{m_Z^2M_{\widetilde{\gamma}}
\cos^2{\beta}}{\bar\mu\,(m_Z^2M_{\widetilde{\gamma}}~\sin 2\beta-
M_1M_2\bar\mu)}~\kappa_i\kappa_j\,,\nonum\\ 
&\equiv& C_{\widetilde S}\; \kap_i\kap_j\,.{\label{treebeitrag}}
\end{eqnarray}
Here $m_Z$ and $M_{1,2}$ denote the mass of the $Z^0$ gauge boson and
the soft supersymmetry breaking, electroweak gaugino mass parameters,
respectively. The photino mass is given by $M_{\widetilde\gamma}\equiv
M_1\cos^2\theta_{W}+ M_2\sin^2\theta_W$, where $\theta_W$ is the
electroweak mixing angle. $|\bar\mu|\equiv\sqrt{\mu^2+\sum_{i}|\kappa
_i| ^2}$. We shall see below that $\sum_{i}|\kappa _i|^2={\cal O}(1\,
{\rm MeV^2})$ and thus to a high precision $\bar\mu\approx\mu$
\cite{Nardi,Grossman,Allanach:2003eb}. For later convenience, we have
introduced the constant $C_{\widetilde S}$ to summarize the dependence
on the supersymmetric parameters.  Within B$_3$-supersymmetry, the
other two neutrinos obtain masses through radiative corrections from
both the bi-linear \cite{Hirsch:2000ef,Diaz:2003as} and the tri-linear
terms in the superpotential \cite{Grossman, davidson}. In the
following, we focus on the radiative corrections due to the tri-linear
terms, since we expect these to dominate for small $\kappa_i$
\cite{Allanach:2003eb}, however realistic neutrino mass models based
only on bi-linear terms have been constructed
\cite{hempfling,Hirsch:2000ef,Diaz:2003as}.

\medskip

There are two distinct radiative contributions to the neutrino masses
from the tri-linear couplings for which the Feynman diagrams are shown
in Fig. ({\ref{neutrinoloops}). One is proportional to $\lam'_{ik\ell}
\lam'_{j  \ell k}$, where a squark and a quark propagate in the loop. A second
is given by a slepton-lepton loop, and is proportional to $\lam_{ik
\ell}\lam_{j\ell k}$.  For the squark-quark loop, the bottom-sbottom
contribution dominates ($k=\ell=3$), since for the down-like quark
masses we have $m_b^2\gg m_s^2,\,m_d^2$.  This results in the mass
matrix\footnote{The extra factor of two arises from two distinct
contributions to the mass which are equal for $k=\ell$
\cite{davidson}.}
\begin{equation}
({m_{\nu}^{d}})_{ij}\approx\frac{2N_c\lam_{i33}^{\prime}\lam_
{j33}^{\prime}}{16\pi^{2}} m_{b}^2A_{b} \frac{f(x_{b})}{M_
{{\tilde b_{2}}}^2}\,,
{\label{squark-quark}}
\end{equation}
where $N_c=3$ is the colour factor and
\begin{equation}
f(x_{b})\equiv\frac{x_b\ln{x_{b}}}{x_{b}-1},\quad {\rm with} \quad x_{b}=
\left(\frac{M_{{\tilde b_{2}}}}{M_{{\tilde b_{1}}}}\right)^2\,.
\end{equation}
$M_{\tilde b_s}$, $s={1,2}$, denote the sbottom mass-eigenstates,
where $M_{{\tilde b_{1}}} < M_{{\tilde b_{2}}}$.  $A_b\equiv
A_b^0-\bar\mu\tan\beta$}, where $A_b^0$ is the tri-linear
soft-super\-symmetry breaking bottom coupling. $i,j=1,2,3$ are
generation indices.  In calculating explicit values for
Eq.~(\ref{squark-quark}), the mixing of the left and right handed
bottom squarks is taken into account but generation mixing is
neglected due to the strict constraints from flavour changing neutral
currents. For later numerical estimates, we find that
\begin{equation}
f(1)=1\,,\qquad f(100)\approx 4.65\,,
\end{equation}
where the latter value corresponds to the fairly extreme value of $M_
{{\tilde b_{2}}}= 10\cdot M_{{\tilde b_{1}}}$. Note, there are no
neutrino mass contributions proportional to $m_bm_s$. The subleading
contribution is proportional to $\lam'_{i22}\lam'_{j22} m_s^2$. This
could in principle be dominant for
$\lam'_{i22}>(m_b/m_s)\lam'_{i33}\approx45\cdot\lam'_{i33}$.  However
due to the relation Eq.~(\ref{mass-ansatz2}), this is not possible in
our ansatz, \textit{cf} Table~\ref{table-LQD}.

\medskip

The contribution from the slepton-lepton loop is analogously given by
($N_c=1$)
\begin{equation}
({m_{\nu}^{e}})_{ij}\approx\frac{\lam_{i33}\lam_{j33}}{8\pi^{2}}
m_{\tau}^2A_{\tau} \frac{f(x_{\tau})}{M_{{\tilde \tau_{2}}}^2}\,.
{\label{slepton-lepton}}
\end{equation}
Here $m_\tau$ is the tau mass and $x_\tau=(M_{\tilde\tau_2}/M_{\tilde
\tau_1})^2$, $M_{\tilde\tau_1}<M_{\tilde\tau_2}$ are the stau masses.
$A_\tau\equiv A_\tau^0-\bar\mu\tan\beta$, where $A_\tau^0$ is the
trilinear soft breaking term for the $\tau$.  Since $\lam_{ijk}$ is
antisymmetric in the first two indices, we must restrict the indices
$i,j=1,2$ in Eq.~(\ref{slepton-lepton}).  For $i=j=3$ the leading
contribution\footnote{In principle, we could also have a contribution
proportional to $\lam_{i22}\lam_{j22}m_\mu^2\propto \ell_3^2 m_\mu^4/
v_d^2$. This can be large since $\ell_3$ is a free parameter in our
ansatz. However, in order to have a comparable contribution, we must
have $\ell_3\approx( m_\tau/mu_\mu)^2\approx 300$. Comparing with
Table~\ref{table-LLE}, we see that the resulting $\lam_{232}$
typically violates the low-energy experimental bounds
\cite{Allanach:1999ic}.} is from the smuon-muon loop, proportional to
$\lam_{322}^2m_\mu^2$.

\medskip

At any given energy scale, the mass parameters $\kap_i$ in
Eq.~(\ref{lep-rpv-superpot}) can be rotated away \cite{rotation}.
Depending on the scale and mechanism of supersymmetry breaking, the
corresponding soft supersymmetry breaking terms will then also vanish
\cite{Allanach:2003eb}.  When this occurs, we have pure tri-linear
$B_3$-models, which have been widely discussed in the literature. The
leading neutrino mass contributions must then arise solely from the
above loop-diagrams. In the following, we shall thus discuss two
models embedded in our ansatz. In Model \textbf{I}, we consider the
case of pure tri-linear interactions, \textit{i.e.} $\kap_i=0$. The
neutrino masses are then solely given through the combined loop
contributions in Eqs.~(\ref{squark-quark}) and (\ref{slepton-lepton}).
In Model \textbf{II}, we consider the more general case with $\kap_i
\not=0$, as well as $\lam,\,\lam'\not=0$. In this case, either
the complete Majorana neutrino mass matrix is given by the sum of all
three contributions, Eqs.~(\ref{treebeitrag}), (\ref{squark-quark})
and (\ref{slepton-lepton}) or one of the loop contributions is absent.
In both models, the neutrino masses and mixing angles are then
obtained upon diagonalization.

\medskip

In the following, we shall first estimate the resulting neutrino
masses in both Models \textbf{I} and \textbf{II} and then determine
the masses and mixing angles by numerically diagonalising the complete
mass matrix.
%
\section{Neutrino Masses from the Simple $\boldsymbol{B_3}$-Model
\label{model}}
\noindent
We now insert our ansatz into the neutrino mass formul{\ae} above and
give an estimate for the $\ell_i,\,\ell_j^\prime$. The tree-level
contribution, Eq.~(\ref{treebeitrag}), is unchanged.  The one-loop
contributions, Eqs.~(\ref{squark-quark}) and (\ref{slepton-lepton}),
are given by
\begin{eqnarray}
({\mathbf{M}_{\nu}^{d}})_{ij}&\approx&\frac{3\ell_{i}^{\prime}\ell_{j}^{\prime}}
{4\pi^{2}}\frac{m_b^4}{\upsilon_d^2}A_b \frac{f(x_{b})}{M_{{b_{2}}}^2}
\equiv3\ell_{i}^{\prime}\ell_{j}^{\prime} C_b
\,,\label{simple-squark-quark}\\
({\mathbf{M}_{\nu}^{e}})_{ij}&\approx&\frac{\ell_{i}\ell_{j}}{4\pi^{2}}
\frac{m_{\tau}^4}{\upsilon_d^2}A_\tau \frac{f(x_{\tau})} {M_{{\tau_{2}}}^2}
\equiv\ell_{i}\ell_{j} C_\tau
\,,\label{simple-slepton-lepton}
\end{eqnarray}
where in the last equation $i,j\not=3$. For later convenience we have
introduced the constants $C_{b,\tau}$.  In order to estimate the order
of magnitude of the $\ell_i,\,\ell_j^\prime $, we shall assume
hierarchical neutrino masses. In Model \textbf{I}, the neutrino masses
are generated alone from the loop corrections and the mass matrix is
then
\begin{eqnarray}
\left[\mathbf{M}_\nu\right]_{ij} = \ell_i\ell_j C_\tau + 3\ell_i'\ell_j' C_b\,,
\end{eqnarray}
which results in two massive neutrinos, since $i,j\not=3$ for
$C_\tau$.  We can obtain a third massive neutrino if we include the
subleading term proportional to $\ell_i\ell_j C_\mu$, for which there
is a contribution for $i,j=3$.

In order to obtain an estimate, we set $\tan\beta=10$ and assume
$A_b^0=A_{\tau}^0=M_{\tilde b,\,\tilde\tau} =m_{soft}=100$ GeV, which
results in $f(x)\ra1$. We then have $C_b\approx 130$ keV and
$C_\tau=4.1$ keV.  The heaviest neutrino is in agreement with the
atmospheric neutrino anomaly, Eq.~(\ref{exp-neut1}), for
\begin{eqnarray}
\ell&\approx&3.5\cdot10^{-3}\,,\quad {\rm or} \quad \;\;\,
\ell^{\prime}\;\approx3.6\cdot 10^{-4}\,,\label{est2a} \\[1mm]
\lam_{i33}&\approx& 3.6\cdot 10^{-4}\,,\quad {\rm or} \quad \lam'_{i33}
\approx 8.7\cdot10^{-5}\,.\label{est2b}
\end{eqnarray}
Correspondingly, we can generate the mass required by the solar
neutrino anomaly by the other term. We obtain $m_{\nu}\approx 0.01$
eV, for
\begin{eqnarray}
\ell&\approx&1.6\cdot10^{-3}\,,\quad {\rm or} \quad
\;\ell^{\prime}\;\;\,\approx1.6\cdot 10^{-4}\,,
\label{est1a} \\
\lam_{i33}&\approx& 1.6\cdot 10^{-4}\,,\quad {\rm or} 
\quad \lam'_{i33}\approx 3.9\cdot 10^{-5}\,, \label{est1b}
\end{eqnarray}
For both the atmospheric and solar anomalies we have used
Eqs.~(\ref{mass-ansatz1}) and (\ref{mass-ansatz2}) to translate back
to the corresponding values for $\lam_{i33}$ and $\lam^\prime_{i33}$.

\medskip

In Model \textbf{II}, the largest neutrino mass is generated by the
tree-level contribution. Taking the trace of Eq.~(\ref{treebeitrag}),
and assuming $M_1\!=\!M_2=\mu=m_{soft}$, we have approximately
\begin{equation}
m_{\nu}\approx\frac{m_Z^2\cos^2\beta\sum_i\kappa_i^2}{m_{soft}^3}\,.
\end{equation}
{}From Eq.~(\ref{exp-neut1}), we then obtain \cite{Nardi}
\begin{equation}  
\sqrt{\sum\kappa_i^2}={\cal O}(1\,{\rm MeV})\,.
\end{equation}
With the same assumptions as in Model \textbf{I}, the radiative contributions
then generate the solar neutrino mass for the values given in
Eqs.~(\ref{est1a}), (\ref{est1b}).
%

\section{Numerical Evaluation of the Neutrino 
Masses \label{numer-eval}}

\subsection{General Outline}

\noindent Next we wish to determine more precisely the parameters of
our ansatz, \textit{i.e.} the $\ell_i,\,\ell^\prime_i,\kap_i$, by
fitting them to the neutrino data. They in turn fix the B$_3$
parameters. In order to learn more about the importance of various
parameters, we consider two cases. In Model \textbf{I}, we consider
the case of pure tri-linear terms. The free parameters are
\begin{eqnarray}
\mathrm{Model\; \mathbf{I}:}\qquad\ell_i,\quad\ell_i'\label{res2}.
\end{eqnarray} 
In Model \textbf{II}, we include $\kap_i\not=0$; the respective free
parameters are given by
\begin{eqnarray}
\mathrm{Model\; \mathbf{II}:}\qquad &\kap_i,&\ell_i,\;\ell_i'\label{res1a}
\end{eqnarray}
Clearly, Model \textbf{II} is the most general case.  

\medskip

The full neutrino mass matrix $(\mathbf{M}_\nu)_{ij}$ is given by the
sum of Eqs. (\ref{treebeitrag}), (\ref{simple-squark-quark}) and
(\ref{simple-slepton-lepton}), where depending on the model, the
coefficient of $C_{\tilde S}$ can be zero.  In the case of Model
\textbf{I}, the real symmetric mass matrix has the form of
\begin{eqnarray}
\mathbf{M}_{\nu}^{\mathbf{I}}&&=\\
&&C_b\left(\begin{array}{ccc}
\ell_1'\ell_1' & \ell_1'\ell_2' & \ell_1'\ell_3'\\
\ell_2'\ell_1' & \ell_2'\ell_2' & \ell_2'\ell_3'\\
\ell_3'\ell_1' & \ell_3'\ell_2' & \ell_3'\ell_3'\\
\end{array}\right)+
C_\tau\left(\begin{array}{ccc}
\ell_1\ell_1 & \ell_1\ell_2 & 0\\
\ell_2\ell_1 & \ell_2\ell_2 & 0\\
0 & 0 & 0\\
\end{array}\right),\nonumber \label{mass-model-I}
\end{eqnarray}
which is a function of five parameters. However, due to the simple
form of the matrices, there are only two non-vanishing neutrino mass
eigenvalues. This is sufficient to explain the atmospheric and solar
neutrino anomalies. In principle, a massless lightest eigenstate can
also lead to an observable effect. If both the massless and the
massive, second lightest neutrino have significant electron-neutrino
admixtures, then the corresponding Kurie plot will have a dip at the
electron energy $Q-m_2$, with $m_2$ being the second lightest neutrino
mass.  The maximal electron energy however will be $Q$, within
experimental uncertainties \cite{Mohapatra:1998rq}. Depending on the
parameter values this could be observable by the KATRIN experiment
\cite{katrin}.

\medskip

In Model~{\bf II}, we obtain three non-zero neutrino masses from the
real symmetric mass matrix,
\begin{eqnarray}
\mathbf{M}_{\nu}^{\mathbf{II}}&=&
C_{\tilde S}\left(\begin{array}{ccc}
\kap_1\kap_1 & \kap_1\kap_2 & \kap_1\kap_3\\
\kap_2\kap_1 & \kap_2\kap_2 & \kap_2\kap_3\\
\kap_3\kap_1 & \kap_3\kap_2 & \kap_3\kap_3\\
\end{array}\right)+\\
&&\hspace{-0.75cm}C_b\left(\begin{array}{ccc}
\ell_1'\ell_1' & \ell_1'\ell_2' & \ell_1'\ell_3'\\
\ell_2'\ell_1' & \ell_2'\ell_2' & \ell_2'\ell_3'\\
\ell_3'\ell_1' & \ell_3'\ell_2' & \ell_3'\ell_3'\\
\end{array}\right)+
C_\tau\left(\begin{array}{ccc}
\ell_1\ell_1 & \ell_1\ell_2 & 0\\
\ell_2\ell_1 & \ell_2\ell_2 & 0\\
0 & 0 & 0\\
\end{array}\right).\nonumber
\end{eqnarray}
which depends on 8 independent parameters. 

\medskip

In the numerical evaluation, the coefficients $C_{\tilde S}$, $C_b$
and $C_{\tau}$ are determined by assuming a BC1 mass spectrum
\cite{Allanach:2006st}, which has a scalar tau LSP. The resulting
neutrino mass matrix $(\mathbf{M}_\nu)_{ij}$ is diagonalized by the
orthogonal rotation matrix $\mathbf{V}$
\begin{equation}
\mathbf{V^TM_{\nu}V}=\left(\begin{array}{ccc}
m_1 & 0 & 0\\
0 & m_2 & 0\\
0 & 0 & m_3\
\end{array}
\right),
\end{equation}
where $m_1\leq m_2\leq m_3$ and $\mathbf{V}$ is given by the
standard parameterization
\begin{eqnarray}
\lefteqn{\mathbf{V}=}\\
&&\hspace{-5mm}\left(\begin{array}{ccc}
c_{12}c_{13} & s_{12}c_{13} & s_{13}\\
-s_{12}c_{23}-c_{12}s_{23}s_{13} & c_{12}c_{23}-s_{12}s_{23}s_{13} 
& s_{23}c_{13}\nonumber\\
s_{12}s_{23}-c_{12}c_{23}s_{13} & -c_{12}s_{23}-s_{12}c_{23}s_{13} 
& c_{23}c_{13}\nonumber
\end{array}\right),
\end{eqnarray}
with $c_{ij}=\cos\theta_{ij}$ and $s_{ij}=\sin\theta_{ij}$. The
complex Dirac phase $\delta_{13}$ and the two Majorana phases
$\alpha_{1/2}$ are omitted. The experimental ranges of the masses and
the mixing parameters $\theta_{12}$ and $\theta_{23}$ are given in
Eqs.~(\ref{exp-neut1})-(\ref{exp-neut4}). In addition, the bound on
the mixing angle $\theta_{13}$ from the CHOOZ experiment
Eq.~(\ref{exp-neut5}) is taken into account. We find it convenient in
presenting our results instead of using the angles as in 
\cite{Gonzalez-Garcia:2004jd,GonzalezGarcia:2007ib}, to use
$\tan^2\theta_{12}$, $\tan^2\theta_{23}$ and $\sin^2\theta_{13}$. The
corresponding $1\sigma$ ranges are given by
\begin{eqnarray}
 \tan^2\theta_{12} &\in& [0.403,0.490]\,, \label{tan-12}\\
\tan^2\theta_{23} &\in& [0.680,1.20]\,, \label{tan-23}\\
\sin^2\theta_{13} &<& 0.0082\,. \label{sin-13}
\end{eqnarray}
In performing the fit, we randomly sample $\log_{10}(\ell_i)$ and
$\log_{10}(\ell_j^\prime)$, thus guaranteeing that we explore the full
hierarchy of couplings. We only consider couplings $\ell_i,\,
\ell_j^\prime\geq 1\cdot10^{-7}$, as smaller couplings have no effect
on the neutrino observables. We furthermore require
\begin{eqnarray}
\ell_i\ell_j \cdot C_\tau &\leq& 0.1\,\mathrm{eV} \\
\ell_i^\prime \ell_j^\prime \cdot C_b &\leq& 0.1\,\mathrm{eV} \\
\kap_i\kap_j \cdot C_S &\leq& 0.1\,\mathrm{eV}\,.
\end{eqnarray}
For given values of the parameters $\ell_i,\,\ell_j^\prime,\,\kap_k$,
the resulting mass matrix is computed. The numerical diagonalization
of the mass matrix yields the mass eigenvalues $m_1$, $m_2$, $m_3$ and
the orthogonal transformation matrix $V$ and thus the mixing angles
$\theta_{12}$, $\theta_{23}$ and $\theta_{13}$.  Afterwards, all
experimental requirements, Eqs.~(\ref{exp-neut1})-(\ref{exp-neut5}),
are applied on the mass eigenvalues and on the mixing angles. We
delete models, which do not fall within the 1$\sigma$ ranges.
Throughout we also assume a hierarchical mass spectrum.  Our results
are summarized in Tables \ref{results-I} and \ref{results-II}, which
we discuss in the next section.

\begin{widetext}

\begin{table}[h!]
\begin{tabular}{|c|c|c|c|c|c|c|c|c|c|c|} 
\hline
 & $\ell_1$ & $l_2$ & $\ell_1^\prime$ & $\ell_2^\prime$ & $\ell_3^\prime$ & $\tan^2
\theta_{12}$ & $\tan^2\theta_{23}$ & $\sin^2\theta_{13}$ & 
$\Delta m_{21}^2$ & $\Delta m_{23}^2$\\
\hline
\hline
$\ell_1$ max & $9.38\cdot 10^{-4}$ & $1.69\cdot 10^{-3}$ & $-1.60\cdot 10^{-7}$ & $4.01\cdot 10^{-4}$ & $-5.22\cdot 10^{-4}$ & 0.48 & 0.99 & 0.0078 & 0.081 & 2.5\\
\hline
$\ell_1$ min & $7.10\cdot 10^{-4}$ & $-1.76\cdot 10^{-3}$ & $9.27\cdot 10^{-5}$ & $4.16\cdot 10^{-4}$ & $-5.09\cdot 10^{-4}$ & 0.42 & 1.10 & 0.0037 & 0.077 & 2.6\\
\hline
$\ell_2$ max & $-7.12\cdot 10^{-4}$ & $1.78\cdot 10^{-3}$ & $-9.42\cdot 10^{-5}$ & $-4.10\cdot 10^{-4}$ & $5.04\cdot 10^{-4}$ & 0.42 & 1.11 & 0.0038 & 0.081 & 2.5\\
\hline
$\ell_2$ min &  $9.23\cdot 10^{-4}$ & $1.58\cdot 10^{-3}$ & $1.82\cdot 10^{-6}$ & $3.83\cdot 10^{-4}$ & $-5.60\cdot 10^{-4}$ & 0.49 & 0.72 & 0.0056 & 0.078 & 2.7\\
\hline
$\ell_1^\prime$ max & $7.10\cdot 10^{-4}$ & $-1.76\cdot 10^{-3}$ & $9.27\cdot 10^{-5}$ & $4.16\cdot 10^{-4}$ & $-5.09\cdot 10^{-4}$ & 0.42 & 1.10 & 0.0037 & 0.077 & 2.6\\
\hline
$\ell_1^\prime$ min & $-9.14\cdot 10^{-4}$ & $-1.67\cdot 10^{-3}$ & $1.00\cdot 10^{-7}$ & $4.11\cdot 10^{-4}$ & $-5.35\cdot 10^{-4}$ & 0.47 & 0.95 & 0.0065 & 0.077 & 2.7\\
\hline
$\ell_2^\prime$ max & $9.01\cdot 10^{-4}$ & $-1.77\cdot 10^{-3}$ & $1.10\cdot 10^{-7}$ & $4.28\cdot 10^{-4}$ & $5.16\cdot 10^{-4}$ & 0.43 & 1.18 & 0.0077 & 0.078 & 2.8\\
\hline
$\ell_2^\prime$ min &$8.73\cdot 10^{-4}$ & $1.62\cdot 10^{-3}$ & $4.00\cdot 10^{-7}$ & $3.61\cdot 10^{-4}$ & $-5.53\cdot 10^{-4}$ & 0.41 & 0.68 & 0.0053 & 0.080 & 2.4\\
\hline
$\ell_3^\prime$ max & $8.69\cdot 10^{-4}$ & $1.61\cdot 10^{-3}$ & $-5.50\cdot 10^{-7}$ & $3.82\cdot 10^{-4}$ & $5.71\cdot 10^{-4}$ & 0.42 & 0.69 & 0.0043 & 0.078 & 2.8\\
\hline
$\ell_3^\prime$ min & $-7.12\cdot 10^{-4}$ & $1.78\cdot 10^{-3}$ & $-9.42\cdot 10^{-5}$ & $-4.10\cdot 10^{-4}$ & $5.04\cdot 10^{-4}$ & 0.42 & 1.11 & 0.0038 & 
0.081 & 2.5\\ \hline
\end{tabular}
\caption{Explicit solutions for Model \textbf {I}, where $\kap_i=0$. 
         We only show the values where one of the $|\ell_i|,\,|\ell_
         j^\prime|$ takes on an extremal absolute value, \textit{i.e.} 
         the highest or lowest value obtained in our fits. Note that 
         due to our sampling constraints, we did not probe values for
         the $\ell_i,\,\ell_j^\prime< 10^{-7}$. In the five columns on
         the far right, we show the resulting neutrino parameters. The
         values for $\Delta m_{21}^2$ and $\Delta m_{23}^2$ are 
         given units $10^{-3}\,\mathrm{eV}^2$. Comparing with Eqs. 
         (\ref{exp-neut1})-(\ref{exp-neut5}), we can see which physical 
         parameter is at its experimentally allowed limit. Thus for 
         example in the first row, where $\ell_1$ is maximal, $\tan^2
         \theta_{12}$ and $\Delta m^2_{21}$ are at the edge of their 
         allowed values. Pushing $\ell_1$ any higher would violate 
         these constraints.}
\label{results-I}
\end{table}

\begin{table}[h!]
\begin{tabular}{|c|c|c|c|c|c|c|c|c|} 
\hline
& $\ell_1$ & $\ell_2$ & $\ell_1^\prime$ & $\ell_2^\prime$ & $\ell_3^\prime$ & $\kappa_1$ & 
$\kappa_2$ & $\kappa_3$ \\ \hline
\hline
$\ell_1$ max & $ 1.08\cdot 10^{-3}$ & $\;-1.17\cdot 10^{-3}\;$ & $1.80\cdot 10^{-6}$ & $-4.39\cdot 10^{-4}$ & $4.57\cdot 10^{-4}$ & $-0.128$ &
 $0.00551$ & $-3.40$\\
\hline
$\ell_1$ min & $1.30\cdot 10^{-7}$ & $1.45\cdot 10^{-5}$ & $-6.50\cdot 10^{-7}$ & $-4.30\cdot 10^{-4}$ & $5.35\cdot 10^{-4}$ & $1.93$ & $2.82$ & $1.09$\\
\hline
$\ell_2$ max & $-8.15\cdot 10^{-4}$ & $1.77\cdot 10^{-3}$ & $4.06\cdot 10^{-5}$ & $4.14\cdot 10^{-4}$ & $-5.01\cdot 10^{-4}$ & $-0.159$ & $-0.541$ & $-0.103$\\
\hline
$\ell_2$ min & $4.50\cdot 10^{-7}$ & $1.20\cdot 10^{-7}$ & $1.42\cdot 10^{-4}$ & $3.22\cdot 10^{-4}$ & $4.10\cdot 10^{-7}$ & $1.02$ & $-4.65$ & $6.11$\\
\hline
$\ell_1^\prime$ max & $\;-8.80\cdot 10^{-7}\;$ & $1.34\cdot 10^{-3}$ & $1.84\cdot 10^{-4}$ & $2.28\cdot 10^{-4}$ & $2.20\cdot 10^{-6}$ & $0.401$ & $\;-4.64\;$ & $6.01$\\
\hline
$\ell_1^\prime$ min & $-4.65\cdot 10^{-6}$ & $9.62\cdot 10^{-6}$ & $2.20\cdot 10^{-7}$ & $-5.28\cdot 10^{-4}$ & $-4.04\cdot 10^{-4}$ & $-2.07$ & $-0.0320$ & $3.72$\\
\hline
$\ell_2^\prime$ max &$-1.50\cdot 10^{-6}$ & $-7.07\cdot 10^{-6}$ & $\;-8.85\cdot 10^{-5}\;$ & $5.31\cdot 10^{-4}$ & $-4.10\cdot 10^{-4}$ & $1.70$ & $0.100$ & $-3.81$\\
\hline
$\ell_2^\prime$ min & $-9.49\cdot 10^{-4}$ & $1.66\cdot 10^{-3}$ & $-9.70\cdot 10^{-6}$ & $1.30\cdot 10^{-7}$ & $1.10\cdot 10^{-4}$ & $0.219$ & $4.92$ & $5.98$\\
\hline
$\ell_3^\prime$ max & $8.82\cdot 10^{-4}$ & $1.64\cdot 10^{-3}$ & $3.60\cdot 10^{-7}$ & $3.91\cdot 10^{-4}$ & $5.55\cdot 10^{-4}$ & $-0.457$ & $0.0604$ 
& $-0.218$\\
\hline
$\ell_3^\prime$ min & $4.50\cdot 10^{-7}$ & $1.20\cdot 10^{-7}$ & $1.42\cdot 10^{-4}$ & $3.22\cdot 10^{-4}$ & $4.10\cdot 10^{-7}$ & $1.02$ & $-4.65$ & $6.11$\\
\hline
$\kappa_1$ max &$1.43\cdot 10^{-4}$ & $-1.30\cdot 10^{-3}$ & $-2.69\cdot 10^{-6}$ & $4.47\cdot 10^{-4}$ & $-4.77\cdot 10^{-4}$ & $2.06$ & $-0.0125$ & $\;-3.03$\\
\hline
$\kappa_1$ min & $-1.87\cdot 10^{-6}$ & $-2.30\cdot 10^{-5}$ & $-1.74\cdot 10^{-4}$ & $-4.23\cdot 10^{-5}$ & $2.99\cdot 10^{-4}$ & $0.0149$ & $-5.90$ & $-5.32$\\
\hline
$\kappa_2$ max & $1.21\cdot 10^{-4}$ & $3.77\cdot 10^{-5}$ & $-1.65\cdot 10^{-4}$ & $-1.04\cdot 10^{-4}$ & $2.27\cdot 10^{-4}$ & $0.139$ & $6.18$ & $5.26$\\
\hline
$\kappa_2$ min & $9.03\cdot 10^{-4}$ & $-1.19\cdot 10^{-3}$ & $-2.15\cdot 10^{-5}$ & $4.56\cdot 10^{-4}$ & $-4.49\cdot 10^{-4}$ & $-0.556$ & $0.0338$ & $3.16$\\
\hline
$\kappa_3$ max& $-7.67\cdot 10^{-4}$ & $-1.23\cdot 10^{-3}$ & $-8.03\cdot 10^{-5}$ & $-1.07\cdot 10^{-4}$ & $1.31\cdot 10^{-4}$ & $\;-0.0348\;$ & $5.00$ & $6.42$\\
\hline
$\kappa_3$ min & $9.00\cdot 10^{-4}$ & $-1.60\cdot 10^{-3}$ & $2.30\cdot 10^{-6}$ & $\;-3.75\cdot 10^{-4}\;$ & $-5.64\cdot 10^{-4}$ & $0.0656$ & $\;-0.0242\;$ & $\;0.0140\;$\\
\hline
\end{tabular}
\caption{The same as Table \ref{results-I} for Model \textbf{II}. Now
we also have the values of $\kap_i$ as well as the extremal values of
$|\kap_i|$. The $\kap_i$ are given in units of $\mathrm{MeV}$.}
\label{results-II}
\end{table}

\end{widetext}

%
%

\begin{table}[h!]
\begin{tabular}{|c|c|c|c|c|c|} 
\hline
&$\tan^2\theta_{12}$ & $\tan^2\theta_{23}$ & $\sin^2\theta_{13}$ & 
$\Delta m_{21}^2$ & $\Delta m_{23}^2$\\
\hline
\hline
$\ell_1$ max & 0.48 & 0.80 & 0.0061 & 0.077 & 2.4\\
\hline
$\ell_1$ min &  0.45 & 0.79 & 0.0012 & 0.081 & 2.5\\
\hline
$\ell_2$ max & 0.41 & 1.19 & 0.0007 & 0.078 & 2.5\\\hline
$\ell_2$ min &  0.44 & 0.93 & 0.0031 & 0.080 & 2.4\\
\hline
$\ell_1^\prime$ max & 0.41 & 1.06 & 0.0003 & 0.079 & 2.4\\
\hline
$\ell_1^\prime$ min &0.47 & 1.04 & 0.0072 & 0.081 & 2.6\\
\hline
$\ell_2^\prime$ max &  0.44 & 1.03 & 0.0009 & 0.080 & 2.7\\
\hline
$\ell_2^\prime$ min & 0.47 & 1.04 & 0.0075 & 0.080 & 2.6\\
\hline
$\ell_3^\prime$ max & 0.44 & 0.78 & 0.0053 & 0.079 & 2.7\\
\hline
$\ell_3^\prime$ min & 0.44 & 0.93 & 0.0031 & 0.080 & 2.4\\
\hline
$\kappa_1$ max & 0.40 & 0.83 & 0.0034 & 0.080 & 2.6\\
\hline
$\kappa_1$ min & 0.47 & 0.81 & 0.0056 & 0.081 & 2.7\\
\hline
$\kappa_2$ max & 0.49 & 1.18 & 0.0001 & 0.078 & 2.6\\
\hline
$\kappa_2$ min & 0.42 & 0.95 & 0.0051 & 0.081 & 2.4\\
\hline
$\kappa_3$ max& 0.44 & 0.74 & 0.0014 & 0.081 & 2.8\\
\hline
$\kappa_3$ min &  0.45 & 0.69 & 0.0055 & 0.080 & 2.6\\
\hline
\end{tabular}
\caption{Model {\bf II} (continued). Again, the values for $\Delta 
         m_{21}^2$ and $\Delta m_{23}^2$ are given units $10^{-3}\,
         \mathrm{eV}^2$.}
\end{table}

\subsection{Discussion of the Results}

\subsubsection{Model \textbf{I}}

In Table \ref{results-I}, we present the fit values for the parameters
$\ell_i,\,\ell_j^\prime$ in Model \textbf{I}. In the five columns on
the right, we also include the resulting neutrino mass and mixing
parameters. Of the large number of solutions we find, we present those
where the parameters $|\ell_i|$ take on extremal values. For example
in the first row of Table \ref{results-I}, $|\ell_1|$ takes on the
largest value we have found. We can now see in the five columns on the
right, that $\tan^2\theta_{12}$ and $\Delta m^2_{21}$, are at the
upper limit of their allowed ranges, Eqs.~ (\ref{exp-neut1}) and
(\ref{tan-12}), respectively. This is as we would expect from
Eq.~(\ref{mass-model-I}), where we see that $\ell_1$ influences the
first two generations. On the other hand, for example in the seventh
row, where $|\ell_2^\prime|$ is maximal, we see that
$\tan^2\theta_{23}$ and $\Delta m^2_{32}$ are at the upper limit of
their allowed ranges.  Similarly, in the second row, where $|\ell_1|$
is minimal, we see that $\Delta m^2_{21}$ is at the \textit{lower} end
of its allowed range. For the eighth row, where $|\ell_2^\prime|$ is
minimal, $\Delta m^2_{32}$ is at the \textit{lower} end of its allowed
range. Overall, we see that $\tan^2\theta_{12}$ is at its upper limit
for [$\ell_1$ max] and also for [$\ell_2$ min]. $\tan^2\theta_{23}$ is
at its upper limit for [$\ell^\prime_2$ max] and at its lower limit
for [$\ell^\prime_2$ min] and [$\ell^\prime_3$ max].
$\sin^2\theta_{13}$ is always well within its limits and thus does not
pose a real constraint on our fit. However, we do predict a value
between 0.003 and the current upper bound. For $\Delta m^2_{21}$, we
are at the upper end of the allowed range for [$\ell_1$ max],
[$\ell_2$ max], and [$\ell^\prime_3$ min]. We are close to the lower
range for [$\ell_1$ min], [$\ell_1^\prime$ max], and [$\ell_1^\prime$
min].  For $\Delta m^2_{23}$, we are at the upper end of the allowed
range for [$\ell_2^\prime$ max] and [$\ell_3^\prime$ max]. Thus we get
the strongest constraints from the allowed mass ranges and from
$\tan^2\theta_{23}$.

\medskip

In the case of Model \textbf{I}, we only have five free parameters.
With these we must fit the two neutrino masses, two mixing angles and
one upper bound. It is thus perhaps not surprising, that except for
$\ell_1^\prime$, the allowed ranges for the five parameters are quite
narrow. $\ell^\prime_{1\,\mathrm{min}}$ is consistent with zero. In
summary, we find from Table \ref{results-I}
\begin{eqnarray}
7.10\cdot 10^{-4}&<|\ell_1| &< 9.38\cdot 10^{-4} \\
1.58\cdot 10^{-3}&<|\ell_2| &< 1.78\cdot 10^{-3} \\
1.00\cdot 10^{-7}&<|\ell_1^\prime| &< 9.27\cdot 10^{-5} \\
3.61\cdot 10^{-4}&<|\ell_2^\prime| &< 4.28\cdot 10^{-4} \\
5.04\cdot 10^{-4}&<|\ell_3^\prime| &< 5.71\cdot 10^{-4} \,.
\end{eqnarray}
We shall employ the central values of these regions in the discussion
of the resulting collider signals, below.

\subsubsection{Model \textbf{II}}

In Table \ref{results-II}, we show the results of our fit to Model
\textbf{II}. We now have a total of eight free-parameters. We vary the
values of the mass mixing parameters in the interval
\begin{equation}
0.01\,\mathrm{MeV}\leq \kap_i \leq
10\,\mathrm{MeV}\,.
\end{equation}
Due to the enhanced freedom, we see that we now have solutions, where
are proportionality constants $\ell_i,\,\ell_j^\prime=\mathcal{O}(10^
{-7})$, which is consistent with zero, in our approach. We see that we
push the upper boundary of $\tan^2\theta_{12}$ for [$\ell_1$ max] and
[$\kap_2$ max] and the lower boundary for [$\kap_1$ max]. For $\tan^2
\theta_{23}$ we push the upper and lower limits for [$\kap_2$ max] and
for [$\kap_3$ min], respectively. From $\sin^2\theta_{13}$, we again
have basically no constraint on our model, beyond those of the other
parameters, \textit{i.e.} we are always well within the CHOOZ bound.

\medskip

We are pushing the upper end of the allowed range of $\Delta m^2_{21}$
for [$\ell_1$ min,] [$\ell_1^\prime$ min], [$\kap_{1,2}$ min], and
[$\kap_3$ max]. For $\Delta m^2_{23}$, we are at the upper end for
[$\ell_2^\prime$ max], [$\ell_3^\prime$ max], and [$\kap_1$ min].  We
are at the \textit{lower} end of $\Delta m^2_{23}$ for [$\ell_1$ max],
[$\ell_2$ min], [$\ell_1^\prime$ max], [$\ell_3^\prime$ min], and
[$\kap_2$ min]. Thus the mass ranges set the strictest limits on our
parameters, the angles are fairly easy to accommodate.

\medskip

Due to the enhanced freedom, we see that any one of our parameters can
consistently be set to zero. This is of particular interest when
trying to extract typical collider signatures, below. In Model 
\textbf{II}, it is thus difficult to discern an identifying
experimental signature.

%
\section{Bounds on the Products of the Parameters 
of the Simple B$_3$-Model} \label{bounds}
%
\noindent Before discussing the consequences of our model, we first consider the
low-energy constraints. Typical bounds on single B$_3$ couplings are
of order 0.1 to 0.01 \cite{Barger:1989rk,Dreiner:1997uz}. However, we
necessarily have multiple couplings in our models and thus must take
into account the bounds on products of couplings
\cite{Davidson:1993qk,Smirnov:1996bg,Allanach:1999ic,Dreiner:2006gu},
which are also typically much stricter, due to lepton flavor violating
effects.  The strictest product bounds ($<10^{-7}$) of Table II in
Ref.~\cite{Allanach:1999ic} are given in Table \ref{tab-bounds}.

\begin{table}[t]
\begin{tabular}{|c|c|c|c|c|}\hline
& $\lam_{122}\lam_{211}^\prime$& 
$\lam_{132}\lam_{311}^\prime$& 
$\lam_{121}\lam_{111}^\prime$& 
$\lam_{231}\lam_{311}^\prime$
\\ \hline
Bound & $4.0\cdot10^{-8}$& $4.0\cdot10^{-8}$& $4.0\cdot10^{-8}$& $4.0\cdot10^{-8}$ 
\\ \hline
\end{tabular}

\vspace{0.2cm}

\begin{tabular}{|c|c|c|c|c|c|}\hline
& 
$\lam_{i12}^\prime\lam_{i21}^\prime$& 
$\lam_{113}^\prime\lam_{131}^\prime$& 
$\lam_{i13}^\prime\lam_{i31}^\prime$& 
$\lam_{1k1}^\prime\lam_{2k1}^\prime$& 
$\lam_{11j}^\prime\lam_{21j}^\prime$
\\ \hline
Bound & 
$10^{-9}$& $3\cdot10^{-8}$& $8\cdot10^{-8}$& $8.0\cdot10^{-8}$& $8.5\cdot10^{-8}$
\\ \hline
\end{tabular}
\caption{Bounds on the products of B$_3$ couplings \cite{Allanach:1999ic}. 
The first four and the last two bounds arise from contributions to the
process $\mu$Ti$\ra e$Ti. The fifth bound arises from contributions to
$\Delta m_K$ and the sixth and seventh from contributions to $\Delta
m_B$. \label{tab-bounds}}
\end{table}

We can now investigate whether our couplings satisfy the bounds in
Table \ref{tab-bounds}. For this we use the values in Tables
\ref{table-LLE} and \ref{table-LQD} and assume $\tan\beta=10$.
We find
\begin{eqnarray}
\lam_{122}\lam_{211}^\prime &=& 2.1\cdot 10^{-6}\, \ell_1\, \ell_2^\prime 
\phantom{\quad j=1}\label{strict1}\\
\lam_{132}\lam_{311}^\prime &=& 0 \\
\lam_{121}\lam_{111}^\prime &=& -9.9\cdot 10^{-9}\, \ell_2\, \ell_1^\prime\\
\lam_{231}\lam_{311}^\prime &=&0 \\
\lam_{i12}^\prime\lam_{i21}^\prime &=& 1.1\cdot 10^{-7}\,(\ell_i^\prime)^2\label{strict2} \\
\lam_{113}^\prime\lam_{131}^\prime &=& 2.7\cdot 10^{-9}\, (\ell_1^\prime)^2
\end{eqnarray}
\begin{eqnarray}
\lam_{i13}^\prime\lam_{i31}^\prime &=& 2.7\cdot 10^{-9}\, (\ell_i^\prime)^2\\
\lam_{1k1}^\prime\lam_{2k1}^\prime &=& 1.2\cdot 10^{-7}\, \ell_1^\prime\,\ell_2^
\prime,\quad k=1 \\
\lam_{11j}^\prime\lam_{21j}^\prime &=& 2.0\cdot 10^{-6}\, \ell_1^\prime\,
\ell_2^\prime,\quad j=2  
\end{eqnarray}
In the last two cases the chosen indices result in the largest
possible value. The strictest bounds result from Eqs.~(\ref{strict1})
and (\ref{strict2}): $\ell_1\ell_2^
\prime<0.02$, and $\ell_i^\prime<0.1$. From Tables \ref{results-I} and \ref{results-II}, we
see that these are always satisfied in our numerical solutions.

\section{Collider Tests\label{collider}}

\noindent An essential feature of B$_3$ neutrino models, is that they
necessarily lead to observable consequences at colliders. Resonant
slepton or squark production requires couplings of order $10^{-3}$ or
larger \cite{resonant}. As can be seen from Tables~\ref{table-LLE} and
\ref{table-LQD} together with the numerical results presented in
Tables \ref{results-I}, \ref{results-II}, this is not possible in our
models.  However, it is well known, that at the LHC squark and gluino
production provide the largest supersymmetric cross sections. This is
independent of whether P$_6$ or B$_3$ is the relevant symmetry. The
produced squarks and gluinos then cascade decay within the detector to
the LSP. In particular, this also holds for the BC benchmark points
\cite{Allanach:2006st}, where $\tilde\tau_1$ is the LSP. In this paper,
we focus on stau-LSP scenarios, as outlined in
Ref.~\cite{Allanach:2006st}. We shall focus on the essential features,
a full phenomenological analysis goes beyond the scope of this paper
and will be presented elsewhere \cite{pheno}. The neutralino LSP case
requires a full treatment of the scalar potential in order to
determine the relevant couplings and masses. This in turn requires
assumptions about the soft-supersymmetry breaking sector, which also
goes well beyond the scope of this paper. We shall consider this
elsewhere \cite{pheno}.

\begin{center}
\begin{figure}[t]
\hspace*{-10mm}
\includegraphics[scale=0.8]{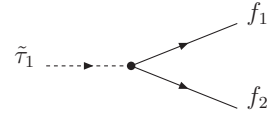}
\caption[Fig.]{\label{staudecay}\footnotesize 
The two-body decay of a scalar particle into two fermions.}
\label{two-body}
\end{figure}
\end{center}

\vspace{-0.6cm}

\subsection{Stau LSP Decays}

As discussed in detail in
Refs.~\cite{Allanach:2003eb,Allanach:2006st}, there are extensive
regions of mSUGRA parameter space, where the scalar tau is the LSP.
The final state collider signals will be determined by the dominant
decays of the stau. The lightest stau, $\tilde\tau_1$, is an admixture
of right and left stau.
\begin{equation}
\tilde\tau_1=\cos\theta_{\tilde\tau}\, \tilde\tau_R + \sin\theta_{\tilde\tau} 
\,\tilde\tau_L
\end{equation}
with $\theta_{\tilde\tau}$ the mixing angle. In
Ref.~\cite{Allanach:2006st}, it was found that in the representative
benchmark points (BC1-BC4) the $\tilde\tau_1$ is dominantly a
right-handed stau with $|\theta_{\tilde\tau}|<0.3$ (in radiants),
\textit{i.e.} the $\tilde\tau_1$-LSP is more than 91\% $\tilde\tau_R$
and $\sin^2\theta_{\tilde\tau}< 0.09$.

\medskip

In our model, we have a wide range of non-zero B$_3$ couplings, where
the corresponding operators couple directly to the stau. The stau can
thus decay via the two-body mode into two spin-1/2 fermions $f_{1,2}$
shown in Fig.~\ref{two-body}. The corresponding partial decay rate is
in given in leading order by \cite{Dreiner:1999qz}
\begin{eqnarray}
\Gamma(\tilde\tau_1\ra f_1 f_2) &=& 
\frac{N_c|\Lam|^2\Theta^2 p_{cm}}{8\pi M_{\tilde\tau_1}^2}
(M_{\tilde\tau_1}^2-m_1^2-m_2^2)\;\;\;\;\;\, \\
&\approx& \frac{N_c|\Lam|^2\Theta^2}{16\pi}\;M_{\tilde\tau_1}\,, 
\label{approx}
\end{eqnarray}
where
\begin{equation}
p_{cm}^2 = \frac{\left[M^2_{\tilde\tau_1}-(m_1+m_2)^2\right]
\left[M^2_{\tilde\tau_1}-(m_1-m_2)^2\right]}{4 M_{\tilde\tau_1}^2}
\,.
\end{equation}
$m_{1,2}$ denote the final state masses. $\Lam$ denotes one of the
following B$_3$ couplings relevant for tree-level stau decay
\begin{equation}
\Lam \in \{\lam_{131},\,\lam_{133},\,\lam_{232},\,\lam_{233},\,
\lam^\prime_{3jk} \}\,,
\end{equation}
and $N_c$ is the colour factor. $N_c=1$ for the decay via the $LL\bar
E$ operators, and $N_c=3$ for the $LQ\bar D$ operators. $\Theta=(\cos
\theta_{\tilde\tau},\sin\theta_{\tilde\tau})$, depending on whether
the $\tilde\tau_1$ couples via the right- or the left-handed stau
component. Eq.~(\ref{approx}) is taken for the case where
$m_{1,2}\ra0$. This is a good approximation for $M_{\tilde\tau_1}<
m_{\mathrm{top}}$, which is the case for all the BC benchmark points.

Given the above decay formula we can now compute the decay rates for
the dominant decay modes using the numerical values in Tables
\ref{table-LLE} and \ref{table-LQD} for the relevant coupling. We
expect the decays where the right-handed stau component couples
directly to dominate, due to the small mixing angle in the stau
sector.  Furthermore, for $\lam'_{333}$ which is a potentially large
coupling the large top quark mass kinematically blocks the decay, for
the stau masses we consider here. We present the results for the
decays in terms of the $\ell_i,\,\ell^\prime_j$ in Table
\ref{stau-decay-rates}.  For completeness, we have included the
couplings involving $\ell_3$, which we have neglected in our neutrino
parameter fits. We see that for substantial decays via the
corresponding operators, we would require, \textit{e.g.} $\ell_3\gg\ell
_{1,2}$.

\medskip

\begin{table}
\begin{tabular}{|c|c|c|}\hline
Operator & Decay Mode & $\;\Gamma(\tilde\tau_1^-\ra
bc)/M_{\tilde\tau_1}\;$ \\ \hline 
$L_1L_3\bar E_1$ & $\;\tilde\tau_1^-\ra e^-\nu_e\;$ & 
$\;1.5\cdot10^{-12}\;\ell_3^2\;$ \\ \hline 
$L_1L_3\bar E_3$ & $\;\tilde\tau_1^-\ra\tau^-\nu_e\;$ &
$\;2.0\cdot10^{-4}\;\ell_1^2\;$ \\ \hline 
$L_1L_3\bar E_3$ &
$\;\tilde\tau_1^-\ra e^-\nu_\tau\;$ & $\;1.8\cdot10^{-4}\;\ell_1^2\;$
\\ \hline 
$L_2L_3\bar E_2$ & $\;\tilde\tau_1^-\ra\mu^-\nu_\mu\;$ &
$\;6.7\cdot10^{-8}\;\ell_3^2\;$ \\ \hline 
$L_2L_3\bar E_3$ &
$\;\tilde\tau_1^-\ra\tau^-\nu_\mu\;$ & $\;2.0\cdot10^{-4}\;\ell_2^2\;$
\\ \hline 
$L_2L_3\bar E_3$ & $\;\tilde\tau_1^-\ra\mu^-\nu_\tau\;$ &
$\;1.8\cdot10^{-4}\;\ell_2^2\;$ \\ \hline 
$\;L_3Q_2\bar D_3\;$ &
$\;\tilde\tau_1^-\ra c\,b\;$ &
$\;\;5.4\cdot10^{-7}\;(\ell_3^\prime)^2\;\;$ \\ \hline $\;L_3Q_2\bar
D_2\;$ & $\;\tilde\tau_1^-\ra c\,s\;$ &
$\;\;1.8\cdot10^{-7}\;(\ell_3^\prime)^2\;\;$ \\ \hline $\;L_3Q_1\bar
D_2\;$ & $\;\tilde\tau_1^-\ra u\,s\;$ &
$\;\;1.1\cdot10^{-8}\;(\ell_3^\prime)^2\;\;$ \\ \hline
\end{tabular}
\caption{Decay modes and partial decay widths of a stau LSP given as a
         function of the relevant $\ell_i,\,\ell^\prime_j$. In the 
         second and the fifth decay modes we have added the two 
         contributions from the doublet and the singlet stau. We have
         set $\theta_{\tilde\tau}=0.3$ (in radiants) as obtained for BC1.}
\label{stau-decay-rates}
\end{table}

\subsubsection{Model I}

If we consider the BC1 benchmark point, we have $M_{\tilde\tau_1}=148
.38$ GeV. Using Tables \ref{results-I}, \ref{results-II} we can then
compute explicit values for the partial widths and the branching
ratios. In Model \textbf{I}, $|\ell_1|\approx 8\cdot 10^{-4}$, $|\ell
_2|\approx2\cdot10^{-3}$ and $|\ell_3^\prime|\approx5\cdot 10^{-4}$.
Using these values, we then obtain the total decay width and lifetime
in Model \textbf{I} at BC1
\begin{eqnarray}
\Gamma(\tilde\tau_1)&=& 260 \,\mathrm{eV}\,,\quad  \tau(\tilde\tau_1)= 
2.5\cdot 10^{-18}\,\mathrm{sec}\,.
\end{eqnarray}
We see that the stau-LSP always decays within the detector. It also
will not lead to a detached vertex. For the branching ratios of the
decay modes in Table \ref{stau-decay-rates}, we obtain 
\begin{eqnarray}
\mathrm{Br}(\tilde\tau_1\ra\tau^-\nu_e) &=& 0.072 \\
\mathrm{Br}(\tilde\tau_1\ra e^-\nu_\tau) &=& 0.065 \\
\mathrm{Br}(\tilde\tau_1\ra\tau^-\nu_\mu) &=& 0.45 \\
\mathrm{Br}(\tilde\tau_1\ra\mu^-\nu_\tau) &=& 0.41 
\end{eqnarray}
The other decay modes are negligible. The branching ratios are
independent of $M_{\tilde\tau_1}$, \textit{i.e.}  in Model~\textbf{I}
they only depend on the $\ell_i,\,\ell_j^\prime$. As the neutrinos are
not visible, we have a combined branching ratio into charged tau
leptons of about 52\%. For squark or gluino pair production, we expect
two stau's in the decay chains. The probability for then having two
charged electrons/muons in the final state is about 23\%. Since the
gluino/squark pair production cross section is very large for
accessible supersymmetric masses, this should lead to an easily
visible signal rate.

\subsubsection{Model \textbf{II}}

In Model \textbf{II}, we have eight free parameters and thus a much
larger freedom. Furthermore for $\kap_3\not=0$ the stau can mix with
the charged Higgs boson, leading to additional decay modes. In order
to compute these properly, we must minimize the full scalar potential.
This is beyond the scope of this paper. We can estimate the stau-Higgs
mixing to be $\kap_3/\mu$. Using the Feynman rules in Fig.~8 of
Ref.~\cite{Gunion:1984yn} for the charged Higgs coupling to the tau
lepton, we then typically find a product of mixing times couplings of
order $10^{-7}$, for $\kap_3=1$ MeV, $\tan\beta=10$ and $\mu=200$
GeV. This would lead to an additional decay $\tilde\tau\ra\tau\nu$.
The couplings to the second generation quarks are another order of
magnitude smaller. These Higgs-mixing couplings are negligible
compared to the direct stau decay couplings, in most cases. However,
in general they must be included. A proper complete treatment will be
given in Ref.~\cite{pheno}. Here it shall suffice to present one
example case from Table \ref{results-II} employing only the direct
decays from Table \ref{stau-decay-rates}. We choose the example, such
that $\ell_{1,2} \ll \ell^\prime_3$, which differs from
Model~\textbf{I}.

\medskip

We consider the case: [$\ell_1^\prime$ min], where $\ell_1=-4.65\cdot
10^{-6}$, $\ell_2=9.62\cdot 10^{-6}$ and $\ell_3^\prime=-4.04\cdot10^
{-4}$. We find for the total decay rate and the lifetime
\begin{eqnarray}
\Gamma(\tilde\tau_1)&=& 2.4\cdot 10^{-2} \,\mathrm{eV}\,,\\
\tau(\tilde\tau_1)&=& 2.7\cdot 10^{-14}\,\mathrm{sec}\,.\;
\end{eqnarray}
We see that the width is now substantially smaller and thus the
lifetime correspondingly larger. For a Lorentz boost $\gamma_L=10$, we
have a decay length of about 100$\mu$m. This is on the boarderline of
visibility for a detached vertex. For the branching ratios we find
\begin{eqnarray}
\mathrm{Br}(\tilde\tau_1\ra\mathrm{hadrons}) &=& 0.73 \\
\mathrm{Br}(\tilde\tau_1\ra\tau^-\nu_e) &=& 0.026 \\
\mathrm{Br}(\tilde\tau_1\ra e^-\nu_\tau) &=& 0.023 \\
\mathrm{Br}(\tilde\tau_1\ra\tau^-\nu_\mu) &=& 0.11 \\
\mathrm{Br}(\tilde\tau_1\ra\mu^-\nu_\tau) &=& 0.10 
\end{eqnarray}
The stau now dominantly decays hadronically. In this specific case, we
can have still a roughly 25\% branching ratio to charged leptons. Or a
probability of roughly 6\% for two charged leptons in the final state.
However, recall that we only scanned couplings down to $10^{-7}$. Thus
we would expect solutions with even smaller $\ell_{1,2}$. In this case
we would have purely hadronic final states and we must resort to the
techniques used in Ref.~\cite{udd}, where the $\bar U\bar D\bar D$
R-parity violating operators were studied.

%
%
\section{Conclusions and Outlook\label{concl}}
\noindent
We have presented a simple ansatz for the B$_3$ Yukawa couplings,
relating them directly to the corresponding Higgs Yukawa couplings via
a small set of parameters $\ell_i,\,\ell^\prime_j$. This results in
simple relations between the B$_3$ couplings presented in Tables
\ref{table-LLE} and \ref{table-LQD}.  We have given estimates of these
parameters in order to obtain the correct neutrino masses and have
then numerically determined the precise values. These are summarised
in Tables \ref{results-I}, \ref{results-II}. We then discussed the
resulting collider signals for the case of a stau LSP. Depending on
the fit values, we have found a wide range for the possible branching
ratios of the stau-LSP. In forthcoming work, we shall give a detailed
investigation of how to disentangle these models at the LHC.

\medskip

%
\centerline{\textbf{\small ACKNOWLEDGMENTS}}

{\color{White}.}\vspace{0.1cm}

\noindent We are grateful to Wilfried Buchm\"uller and Reinhold 
R\"uckl for initiating the discussion on this ansatz and for providing
us with their private notes. We thank Markus Bernhardt for valuable
help and discussions on the B$_3$ version of {\tt{SOFTSUSY}}. We thank
Christoph Luhn, Ulrich Langenfeld and Sebastian Grab for helpful
discussions and Alejandro Ibarra and Laura Covi for valuable comments.
%
%

\end{document}